\input phyzzx
\newcount\mongocount
\mongocount=1
\def\Figure#1#2#3{
% \boxit{
      \vbox to #3in{\hsize=#2in
        \vfil
%\special{ps::[begin]
%          save 10 dict begin /Figure exch def
%          currentpoint translate
%          /showpage {} def
%        }
%        \special{ps: plotfile #1}
         \includegraphics{#1}
%        \special{ps::[end]
%        clear Figure end restore
%        }
    }
% }
}
\def\figcap#1#2{
\vtop{\tenpoint\singlespace
\hsize=#1in\smallskip\noindent Figure\ \ \the\mongocount.\ \  #2
\global\advance\mongocount by 1\bigskip}}
\def\mongofigure#1#2#3#4#5{\centerline{\Figure{#1}{#2}{#3}
\figcap{#4}{#5}}}

\hoffset=0.375in
\overfullrule=0pt

\def\au{{\rm AU}}

\def\dol{{d_{\rm ol}}}
\def\dls{{d_{\rm ls}}}
\def\dos{{d_{\rm os}}}
\def\max{{\rm max}}

\def\kpc{{\rm kpc}}

\def\kms{{\rm km}\,{\rm s}^{-1}}
\def\bv{{\bf v}}
\def\bu{{\bf u}}
\twelvepoint
\font\bigfont=cmr17
\centerline{\bigfont Microlens Parallax Asymmetries}
\smallskip
\centerline{\bigfont Toward the Large Magellanic Cloud}
\bigskip
\centerline{{\bf Andrew Gould}\footnote{1}{Alfred P.\ Sloan Foundation Fellow}}
\smallskip
\centerline{Dept of Astronomy, Ohio State University, Columbus, OH 43210}
\smallskip
\centerline{gould@astronomy.ohio-state.edu}
\bigskip
\centerline{\bf Abstract}
\singlespace 
%\doublespace

	If the microlensing events now being detected toward the Large
Magellanic Cloud (LMC) are due to lenses in the Milky Way halo, then the 
events should typically have asymmetries of order 1\% due to parallax 
from the reflex motion of the Earth.  By contrast, if the lenses are in the 
LMC, the parallax effects should be negligible.  A ground-based search for 
such  parallax asymmetries would therefore clarify the location of the lenses.
A modest effort (2 hours per night on a 1 m telescope) could measure
15 parallax asymmetries over 5 years and so marginally discriminate between
the halo and the LMC as the source of the lenses.  A dedicated 1 m telescope
would approximately double the number of measurements and would therefore
clearly distinguish between the alternatives.  However, compared to
satellite parallaxes, the information
extracted from ground-based parallaxes is substantially less useful 
for understanding the nature of the halo
lenses (if that is what they are).  The backgrounds of asymmetries due to
binary-source and binary-lens events are estimated to be approximately 
7\% and 12\% respectively.  These complicate the interpretation of detected
parallax asymmetries, but not critically.

\bigskip
Subject Headings: dark matter -- Galaxy: halo -- gravitational lensing 
-- Magellanic Clouds

\endpage
%\normalspace
\chapter{Introduction}

	 If the 8 microlensing events observed by the MACHO collaboration
toward the Large Magellanic Cloud (LMC) are due
primarily to lenses in a standard Milky Way halo, then these lenses comprise
about half of the halo mass within 50 kpc and have typical masses
$M\sim 0.4\,M_\odot$ (Alcock et al.\ 1997a).  
Such objects could not be composed of hydrogen because
they would burn, and the population would thus be easily detected in star 
counts.  The mass scale is roughly consistent with that of white dwarfs, but
to date there are no plausible schemes for making enough white dwarfs
to account for the microlensing events without also generating various
easily observed effects such as the overproduction of metals.

	A number of alternative suggestions have therefore been advanced
for the origin of these puzzling events.  Sahu (1994) and Wu (1994)
proposed that the events are due to LMC lenses.  Zhao (1998) suggested that
they lie in a dwarf (possibly tidally disrupted) galaxy along the line
of sight to the LMC, and Zartisky \& Lin (1997) claim to have detected such
a structure.  Evans et al.\ (1998) suggested that the lenses could lie in
a warped and flared Milky Way disk.  A variety of arguments have been
advanced against these alternatives (Gould 1995a; Beaulieu \& Sackett 1998;
Alcock et al.\ 1997a,b; Gould 1998; Bennett 1998) and the nature of the 
events remains unresolved.

	The fundamental problem is that microlensing events are generally
described by a five parameter light curve (Paczy\'nski 1986),
$$ F(t;t_0,\beta,t_e,F_0,B) = F_0 A[u(t;t_0,\beta,t_e)] + B,\eqn\foft$$
where $A(u) = (u^2+2)/[u(u^2+4)^{1/2}]$, 
$u(t)$ is the projected separation of the source and lens in units
of the Einstein radius, $r_e$,
$$u(t;t_0,\beta,t_e) = \biggl[\beta^2 + \biggl({t-t_0\over t_e}\biggr)^2
\biggr]^{1/2},\eqn\xoft$$
$t_0$ is the time of closest approach, $\beta$ is the impact parameter in
units of $r_e$, $t_e$ is the characteristic time of the event, $F_0$ is
the unmagnified flux of the source star, and $B$ is any background light in
the source aperture that is unaffected by the lensing event.  Of the five
parameters in equation \foft, only $t_e$ is in any way related to the
physical characteristics of the lens.  Moreover, this relation is rather
indirect,
$$t_e = {r_e\over |\bv|}.\eqn\tedef$$
where
$$r_e = \biggl({4 G M \dol\dls\over c^2 \dos}\biggr)^{1/2},\eqn\redef$$
is the Einstein radius, $M$ is the mass of the lens, $\dol,$ $\dls,$ and
$\dos$ are the distances between the observer, source, and lens, 
and $\bv$ is the transverse velocity of the lens relative to the 
observer-source 
line of sight.  Thus, the experiment does not directly measure the mass,
distance, or speed separately.  The mass can be estimated only statistically
and then only given models of the distance and velocity distributions of
the lenses and sources.

	The most straight forward way to determine the location of the
lenses would be to launch a parallax satellite into solar orbit 
(Refsdal 1966; Gould 1994,1995b; Boutreux \& Gould 1996; Gaudi \& Gould 1997a;
Markovi\'c 1998).
Since the Earth-satellite separation ($\sim \au$) is of order the size of
the Einstein radius projected onto the plane of the observer, $\tilde r_e$,
the event parameters as seen from the Earth $(t_0,\beta,t_e)$ will be
substantially different from those seen by the satellite $(t_0',\beta',t_e')$.
From the difference, one can infer the projected velocity $\tilde \bv$ or,
equivalently, the projected Einstein radius, $\tilde r_e$, and the direction of
relative motion, $\phi$,
$$\tilde \bv = {\dos\over \dls}\bv,\qquad \tilde r_e = \tilde v t_e.
\eqn\tildedefs$$  
For disk, halo, and LMC lenses, the projected velocities
are typically 50, 300, and more than $1000\, \kms$.  Hence, parallaxes would
allow one to identify the population to which the lenses belong on an almost
case by case basis.  Even lenses in the Evans et al.\ (1998) warped and flared
disk could be distinguished from halo lenses (which have similar distances
and speeds) because the former would be streaming in the direction of the
disk while the latter would be approximately randomly distributed in $\phi$.

	Since the launching of a satellite is a long-term and expensive
undertaking, a number of other ideas have been explored for resolving the
nature of the lenses.  Binary lenses have a caustic in their magnification
structures.  If this caustic transits the source, then the proper motion
of the lens $\mu = \theta_*/t_*$ can be determined from the radius
crossing time, 
$t_*$, and the angular radius of the source, $\theta_*$ (known from Stefan's
Law).  The proper motion $\mu = v/\dol$ is more than an order of magnitude
larger for halo lenses than for LMC lenses, so these two populations
could be easily distinguished.  In fact, the proper motion of one LMC event
has been tentatively measured (Alcock et al.\ 1997a) and appears to be in
the LMC.  However, binary lenses may not be representative of the lens
population as a whole.  For example, it could be that binary lenses occur
only in the LMC and that no halo lenses are binaries.  Thus, 
no definitive conclusion can be drawn from 
measuring binary-lens proper motions.  Proper motions can
also be obtained for some binary-source events (Han \& Gould 1997).  Unlike
the binary-lens events, binary-source events are not intrinsically biased.
However, they are relatively rare and to date none have been 
unambiguously detected.

	Another possible approach is ground-based parallaxes.  The motion
of the Earth causes the projected source-lens separation to deviate from
the straight line reflected in 
equation \xoft\ (Gould 1992; Gould, Miralda-Escud\'e, \&
Bahcall 1994; Buchalter \& Kamionkowski 1997) and hence causes the light 
curve to deviate from equation \foft.  One measures the same parameters as
with satellite parallaxes, $\tilde \bv$, or equivalently $\tilde r_e$ and 
$\phi$.
There is one published parallax for an event seen toward the Galactic
bulge (Alcock et al.\ 1995) and several more which have not yet been published
(Bennett et al.\ 1997).  These events are typically very
long, $t_e\gsim 90\,$days.  The velocity of the Earth changes substantially
during the course of a long event, and this is what allows the parallax to be
measured.  Most events seen toward the LMC are much shorter, 
$t_e\sim 40\,$days.  Moreover, the LMC
source stars are typically much fainter than the bulge sources, making it
more difficult to acquire the accurate photometry required to detect the 
subtle parallax effect.  However, the
first event seen toward the SMC (Alcock et al.\ 1997c; Palanque-Delabrouille
et al.\ 1998) has $t_e\sim 120\,$days and has an exceptionally bright
source ($V\sim 17$).  These characteristics allowed Palanque-Delabrouille
et al.\ (1998) to put a lower limit on the projected speed
$\tilde v>270\,\kms$.

	Gould et al.\ (1994) showed that even
when the events are too short to allow one to make 
a complete parallax determination (and so measure $\tilde \bv$), 
parallax effects can nonetheless induce 
deviations of the light curves relative to equations \foft\ and \xoft.
It might then be possible to detect the 
``parallax asymmetry'' of some of these events.  
Measurement of this asymmetry yields the parameter combination 
$K_\parallel=\cos\phi/\tilde v$ where $\phi$ is the angle between
$\bf v$ and the
Earth-Sun separation vector. 
Of course, it would be much better to measure $\tilde v$ and $\phi$ 
separately.  In particular, if $K_\parallel$ were found to be consistent
with zero, one would not know whether $\tilde v$ was very large (indicating
that the lens was almost certainly in the LMC) or $|\cos\phi|$ just happened
to be very small.  However, if $K_\parallel$ 
were measured for a sample of events, one might be able to determine 
statistically 
whether most of the lenses were in the LMC (very few detections of 
$K_\parallel$) or
whether many were in the Milky Way halo or disk (many detections of 
$K_\parallel$).

	At the time that Gould et al.\ (1994) analyzed parallax asymmetries,
the only event seen toward the LMC had a time scale $t_e\sim 17\,$days.  If
the lens were in the halo, the asymmetry in the light curve would be
substantially smaller than 1\%, probably too small to detect reliably.  Thus, 
Gould et al.\ (1994) considered parallax asymmetries primarily as a method
to distinguish between halo lenses and disk lenses, the latter generating
much larger asymmetries.  However, T.\ Axelrod (1997, private communication) 
has pointed out that since the typical time scale of the observed events is
now $t_e\sim 40\,$days, the asymmetries expected for halo lenses are
typically much larger.

	Here I elaborate on the work of Gould et al.\ (1994) to develop
an analytic framework within which one can assess the detectability of
parallax asymmetries.  I also identify some
backgrounds which may complicate the interpretation of any detections that
are made.

\chapter{Parallax Asymmetries}

	In the limit of short events, the acceleration of the Earth may
be regarded as constant over the duration of the event.  Equation \foft\
then remains valid provided that equation \xoft\ is replaced by
(Gould et al.\ 1994)
$$u(t) = \biggl[\biggl[\xi\biggl({t-t_0\over t_e}\biggr)\biggr]^2+
\beta^2\biggr]^{1/2},\qquad  \xi(y) 
= y + {1\over 2}\gamma y^2,\eqn\xpoft$$
where
$$\gamma = {v_\oplus\over \tilde v}\,{t_e\over {\rm yr}/2\pi}\cos\phi,
\eqn\gammadef$$
$v_\oplus\sim 30\,\kms$ is the speed of the Earth, and where I have made
use of the fact that the LMC is approximately at the ecliptic pole.
Note that for typical observed LMC events ($t_e\sim 40\,$days) and for
projected velocities that are typical of the halo ($\tilde v\sim 275\,\kms$),
the parameter $\gamma\sim 0.08\cos\phi$ is small compared to unity.  
Also note
that at $t_e\sim 40\,$ days, the assumption that the Earth's acceleration
vector is constant is beginning to break down.  However,
for purposes of estimating the detectability of parallax effects, this
approximation is adequate.

\FIG\one{
Asymmetry functions $W$ and $\tau$ plotted against the impact
parameter $\beta$.  The maximum fractional deviation of the light curve
from its standard Paczy\'nski (1986) form is given by
$(\Delta \ln F)_\max = W\gamma$, where $\gamma$ is the asymmetry parameter
defined by eq.\ \gammadef.  This maximum deviation occurs at times
$t= t_0 \pm \tau t_e$.
}

\topinsert
\mongofigure{ps.fig1x_parc}{6.4}{6.0}{6.4}{
Asymmetry functions $W$ and $\tau$ plotted against the impact
parameter $\beta$.  The maximum fractional deviation of the light curve
from its standard Paczy\'nski (1986) form is given by
$(\Delta \ln F)_\max = W\gamma$, where $\gamma$ is the asymmetry parameter
defined by eq.\ \gammadef.  This maximum deviation occurs at times
$t= t_0 \pm \tau t_e$.
}
\endinsert

	Equation \xpoft\ introduces an asymmetry into the light curve.
The lens gradually speeds up (or slows down) during the course of the event,
so the rise time of the event is longer (or shorter) than the fall time.
The first questions to address are: 
what is the maximal fractional deviation of
the parallax-affected light curve from the unperturbed light curve 
($\gamma=0$), and at what time does this maximum deviation occur?  I find
$$(\Delta \ln F)_\max = \gamma W(\beta), \qquad
t_\max = t_0 \pm\tau(\beta) t_e,
\eqn\dellna$$
where $W(\beta)$ is shown as a bold curve, and $\tau(\beta)$ is shown as a
solid curve in Figure \one.  In making this evaluation, I have used equation
\foft\ and assumed
$B/F_0\ll 1$.  Obviously, in the case of significant blending, the maximum
deviation will be smaller.

	According to Figure \one, the maximum deviation occurs about 
one Einstein crossing
time from the peak of the event, more or less independent of the impact
parameter.  This will have importance in understanding the detectability
of the events.  Note that $W$ also depends only weakly on $\beta$.
The typical value is $W\sim 0.2$.  This implies that the maximum deviation
for a typical halo event is $(\Delta\ln F)_\max\sim 0.2\gamma\sim
1.5\%\cos\phi$.  This is certainly well below the photometric precision of
the microlensing surveys that are searching for events.  However, 
1\% photometry could be achieved by aggressive follow-up
observations even for the typical sources which are faint $(V\sim 20)$ and
in crowded fields.

\FIG\two{
Normalized error, $S$, for the measurement of the asymmetry parameter, 
$\gamma$, plotted as a function of the impact parameter, $\beta$.  The
actual error is given by $\sigma_\gamma = S\sigma_0/(N t_e/\rm day)^{1/2}$
where $\sigma_0$ is the factional flux error for an individual measurement,
and $(N t_e/\rm day)$ is the number of measurements per Einstein crossing
time.  The {\it bold curve} assumes that measurements begin when the source
enters the Einstein ring ($u_i=1.0$) and the {\it solid curve} assumes that
they begin at $u_i=1.5$.  In both cases, the observations are assumed to
end at $t=t_0 + 1.5 t_e$.
}

\topinsert
\mongofigure{ps.fig2x_parc}{6.4}{6.0}{6.4}{
Normalized error, $S$, for the measurement of the asymmetry parameter, 
$\gamma$, plotted as a function of the impact parameter, $\beta$.  The
actual error is given by $\sigma_\gamma = S\sigma_0/(N t_e/\rm day)^{1/2}$
where $\sigma_0$ is the factional flux error for an individual measurement,
and $(N t_e/\rm day)$ is the number of measurements per Einstein crossing
time.  The {\it bold curve} assumes that measurements begin when the source
enters the Einstein ring ($u_i=1.0$) and the {\it solid curve} assumes that
they begin at $u_i=1.5$.  In both cases, the observations are assumed to
end at $t=t_0 + 1.5 t_e$.
}
\endinsert

	The next question is then: with what precision can $\gamma$ be measured
assuming such a program is undertaken?  Of course, $\gamma$
must be measured simultaneously with the other five parameters shown in
equations \foft\ and \xpoft.  Employing standard techniques (e.g., 
Gould \& Welch 1996), I evaluate the covariance matrix $c_{i j}$ of the
six parameters $a_i$, by considering a series of measurements at times $t_k$,
and with errors $\sigma_k$,
$$c = b^{-1},\qquad b_{i j} = \sum_k \sigma_k^{-2}
{\partial F(t_k)\over\partial a_i}\,{\partial F(t_k)\over\partial a_j}.
\eqn\bijdef$$
After taking the derivatives $\partial F(t_k)/\partial a_i$, I evaluate them
assuming $B=\gamma=0$.  I assume that the errors are
$\sigma_k = \sigma_0 A(t_k)$.  That is, they are limited by systematics, not
statistics.  I assume that these intensive observations are triggered when
the event enters the Einstein ring ($u_i=1$, $A(u_i)=1.34$) and end at
$t = t_0 + 1.5 t_e$, and that they are carried on uniformly at a rate
$N$ per day in the interval.  I then find an uncertainty in the determination 
of $\gamma$,
$$\sigma_\gamma = {\sigma_0\over (N t_e/\rm day)^{1/2}}S(\beta),
\eqn\sigmagamma$$
where $S(\beta)$ is shown as a bold curve in Figure \two.  Combining this
equation with equation \gammadef\
gives the signal-to-noise ratio (S/N),
$$%{\rm S\over N}  =
{\gamma\over\sigma_\gamma} = 
4.2\,{N}^{1/2}
\biggl({\sigma_0\over 0.01}\biggr)^{-1}
\biggl({S\over 8}\biggr)^{-1}
\biggl({\tilde v\over 275\,\kms}\biggr)^{-1}
\biggl({t_e\over 40\,\rm days}\biggr)^{3/2}\,
{|\cos\phi|\over 0.7}.
\eqn\sneval$$

	To detect the asymmetry from a typical halo event with
$t_e\sim 40\,$days at the $4\,\sigma$
level therefore requires about one observations per day, 
each with $\sigma_0\sim 1\%$ accuracy.  Assuming that 40\% of
the observing time is lost to the Moon and poor weather, this implies
$\sim 1.7$ observations per clear and dark or gray night over the course of
about 3 months.

\chapter{Realistic Observing Requirements}

	A 1 m telescope with reasonably good throughput detects 
$25~e\ \rm s^{-1}$ from a $V=20$ star.  (The use of broader filters is not
advisable for reasons to be discussed below, and even standard filters may
be too broad.)\ \ Thus, to obtain 1\% photometry on a $V=20$ star requires
a minimum exposure of 400 s on a 1 m telescope.  If one assumes a mean 
sky of $V=21\,\rm mag\,arcsec^{-2}$ and $1''$ seeing, the exposure time is
increased to 900 s.  The fields are generally crowded which ordinarily
would imply longer exposure times to acquire the same S/N.  However, it
is not completely clear that substantially longer exposures would be 
required.  According to equation \foft, the parameters are estimated by
first subtracting out a time-invariant background, $B$.  This means that
only pixel lensing (difference-image photometry) is required,
not measurement of the absolute flux.  Several groups are working on 
developing pixel lensing techniques (Tomaney \& Crotts 1996; Ansari et al.\
1997b; Melchior et al.\ 1998; Alard \& Lupton 1998; Tomaney 1998).  
While none have yet
reached the photon limit on such deep exposures, several groups have come
within a factor of two.  For definiteness, I assume a 50\% penalty for 
crowding, add 60 s for overhead,
and arrive at a total cycle time of 1400 s.
This means that 1 hour per night is required to follow each $V=20$ event.

	The total number of events to be followed depends of course on the
vigor of the microlensing search.  To be concrete, I use as a model the
first two years of data from MACHO (Alcock et al.\ 1997a).  A total of 12
events were, one way or another, selected as proto-candidates.  Many of these
were not recognized until after the event had peaked, but I will assume that
better triggers would have permitted all to be detected as they entered the
Einstein ring.  Events 2, 3, 11, and 12 were eventually rejected as candidates
for various reasons.  However, since events must be following beginning well
before their peak, these proto-candidates would have had to have been 
monitored
in any program designed to detect parallax asymmetries. Indeed, a number
of other events would have probably been followed as well, at least for a 
while.  

	Of these 12 proto-candidates, how many would yield parallax asymmetry 
measurements or useful upper limits?   Only eight of the 12 were eventually
deemed candidates.  Of these, two of the sources (5 and 7) were $V\sim 21$ 
stars.  The S/N for these events would be a factor $\sim 2$ smaller than for
the fiducial $V=20$ star calculated above and hence inadequate
to detect an asymmetry unless the observation program was substantially
more aggressive than the one envisaged above.  Of course, this would have
been recognized immediately, leading to a decision either to ignore these
candidates or to focus more resources on them.  I will assume the latter.
Two other candidates (4 and 9) 
are nominally fairly bright ($V\sim 20$ and $V\sim 19.5$) but this is only
because they are blends.  Both source stars are actually $V>21$.  In this
case, the faintness of the true source would not have been recognized until
the followup was well underway.  If this was recognized sufficiently early,
more telescope resources could be applied to these events or they could have
been dropped from the followup program.  I will assume that these events would
have been recognized as severe blends too late to acquire adequate S/N.
In summary, four (1, 6, 8, and 10) of the eight candidates would yield parallax
measurements under the fiducial observing schedule, two (5 and 7) would yield
parallaxes with four times the normal observing time, and the remaining two 
(4 and 9) would not produce useful information.  Here I have not
considered the time scales, binarity, or other known facts about the events,
in order to focus on the distributions of source brightness and blending.

	I allow for four times longer observation for two of the events and
therefore estimate that it would be necessary to monitor 12 events,
each for an average of 90 minutes per clear dark or gray night
over three months in order to make 6 useful measurements.  This estimate
may seem too conservative since, as I have indicated above, a number of the
events would be dropped from the program prior to the full three months of
observation.  However, as also indicated above, an aggressive program would 
need to monitor some promising events that ultimately fail to be
selected as candidates.  Moreover, as I discuss in \S\ 4, each event may 
require additional observations beyond the 3 month interval in order control
various backgrounds.  I therefore believe that this estimate of 50\% efficiency
is realistic.

	To monitor 6 events per year would require an average of 2.25 hours
per night.  This is of order one quarter of the combined telescope time 
dedicated 
to followup of bulge microlensing events by the PLANET (Albrow et al.\ 
1996,1998) and GMAN (Alcock et al.\ 1996,1997d) collaborations.
I will therefore consider both what could be accomplished assuming that 6
events were followed per year and what could be done with a dedicated 1 m
telescope (assuming that the initial
search strategy were optimized to allow maximal followup).  If the initial
search could provide five events in each of the spring and summer and three
in each of the fall and winter, then a single 1 m telescope could devote
approximately 90 minutes per night to each of them.  These additional events
would have to come from the outlying areas of the LMC.  That is, a 
substantially larger area would have to be searched even during winter
when conditions are unfavorable so as to provide events to monitor during
the spring.
MACHO actually monitored about twice as many stars as they
analyzed in Alcock et al.\ (1997a), the remainder being in more outlying
areas of the LMC.  The EROS collaboration (Ansari 1997a) has recently begun
a search over a still larger area employing a $1\,\rm deg^2$ camera.  Thus,
one may expect an increase in the rate of candidates, especially if the lenses
lie in the halo and not the LMC.  The candidates from the outlying fields
should be less crowded and thus easier to measure.  However, for the same
reason they should also be systematically fainter.  For example, EROS's
exposures in their outermost fields are 900 s compared to 180 s in their
central fields.  Thus it is not clear that the initial searches could provide
fully 16 candidates per year that were bright enough for followup and
on the required schedule.  For definiteness, I estimate that 12 suitable
candidates are found per year (i.e., double the current rate) of which 
1/2 yield useful parallax asymmetry
measurements.  Thus, over 5 years, one could obtain about 15 useful 
measurements with one quarter access to a 1 m telescope and 30 useful 
measurements with dedicated access.

	A possible route to increasing the efficiency of the followup 
telescope would be to initiate the observations at $u_i=1.5$, [$A(u_i)=1.13$] 
rather than at $u_i=1$.  This would be substantially more difficult and
would risk more false detections but, as Figure \two\ shows, the improvement
in S/N would be dramatic, especially at higher impact parameters.  One
reason for this improvement is shown in Figure \one: for higher $\beta$,
the first maximum deviation occurs at $t\lsim t_0- t_e$ which is missed 
if followup begins when the source enters the Einstein ring.  A second
reason is that for higher $\beta$, the peak time of the event ($t_0$) is rather
poorly determined if followup is delayed until the event enters the
Einstein ring.  The asymmetry parameter $\gamma$ is highly correlated with
$t_0$ because both $\partial F/\partial t_0$ and $\partial F/\partial\gamma$
are odd functions of $(t-t_0)$.  See equation \bijdef.

\chapter{Backgrounds}

	The expected amplitude of the parallax asymmetry is only $\sim 1\%$.
There are several backgrounds, both astrophysical and geophysical, that might
produce an asymmetry at this level.  These backgrounds must be eliminated
to the extent possible and accounted for in the analysis to the extent
that they cannot be eliminated.  I identify four such backgrounds.

	First, the sources might be variable.  It is 
straightforward to test various LMC populations for this level of variability
simply by measuring the flux from other stars in the monitoring images.  If
$\sim 1\%$ variability on $\sim 3$ month time scales is not extremely rare, 
then it will be necessary to continue to monitor the source after the end
of the event to determine if it is such a variable.  This is, of course,
also straight forward but it does require substantial additional telescope
time.

	Second, differential refraction could cause a seasonal variation
in the flux recovered from the source stars, since the stars will typically
be observed at higher airmass in the winter than the summer.  If such an
effect were present, it would produce a steady gradient in the recovered
flux over the $\sim 3$ months between the times of maximum expected
asymmetrical deviation ($t\sim t_0 \pm t_e$) and thus give the impression of
an asymmetry when there was none.  The recovered
flux would change as differential refraction pushed background stars of
different color closer to (or farther from) the source star.  Thus, this
effect could mimic asymmetries of either sign.  Again, it would be 
straight forward to test for the general presence of such an effect by 
monitoring other stars in the field.  If $\sim 1\%$ amplitudes were
not rare, one could as before measure the size of the effect for the source 
star by continuing the monitoring after the event was over.  Again, the only
cost would be telescope time.  It would be possible to eliminate this effect
altogether by using sufficiently narrow filters.  However, in view of the
severe S/N requirements, this approach would be self-defeating.  Since the
effect worsens with the width of the filters, very wide filters are probably
not a good route to higher S/N.

	Third, binary source events could be asymmetric (Griest \& Hu 1992)
and so could mimic parallax asymmetries.  Consider for
example a binary whose primary is lensed and whose secondary is 5\% as bright
as the primary.  The secondary lies at $\Delta \bu = (1,1)$
where the first component represents the separation along the direction of
motion relative to the lens, and second represents the separation along
 the perpendicular direction.  Suppose
that the impact parameter of the event was $\beta=0.4$ and the lens lay on
the opposite side of the primary source from its companion.  Then in units
of the primary flux, the total flux would be 1.344 at $t=t_0-t_e$ and 1.386
at $t=t_0+t_e$.  That is the binary would mimic a light curve with an
asymmetry of 1.5\%.  Of course, the detailed structure of the binary-source
light curve would be quite different from one produced by parallax asymmetry,
but since according to equation \sneval\ the asymmetry is only just barely
being detected, in most cases one could not distinguish between the two.

	What fraction of events will have asymmetries due to binaries?  Only
binaries with magnitude differences $\Delta V\lsim 4$ and so mass ratios 
$q\gsim 0.45$ (Henry \& McCarthy 1993) can contribute.  Other
companions contribute too little flux to make a detectable asymmetry.  
Griest \& Hu (1992) have compiled data on binaries from various sources and 
report that $\sim 40\%$ of F stars and $\sim 60\%$ of A stars satisfy this
criterion.  Similarly, only binaries with source-lens
separations $|\Delta \bu|\lsim 3$ 
can contribute, since beyond this radius the flux
from the secondary is nearly constant with time.  However, about
half of the binaries within this radius generate light curves that differ 
substantially from parallax
asymmetries.  For example, if the binary is aligned along the $y$ axis,
$\Delta \bu = (0,\Delta u_y)$, then the light curve distortion will be
symmetric.  If it lies along the $x$ axis, the distortion will be double 
peaked.  The fraction of binaries satisfying $|\Delta \bu|\lsim 3$  depends on
what physical separation corresponds to an Einstein radius.  As a practical
matter, however, the dependence is rather weak.  If for example the lenses
are $0.5\,M_\odot$ stars in the LMC, with lens-source
separations $\dls=3\,\kpc$, then
an Einstein radius corresponds to $(\dos/\dol)r_e\sim 3.5\,\au$.  If they
are $0.5\,M_\odot$ objects in the halo with $\dol\sim 10\,\kpc$, then
$(\dos/\dol)r_e\sim 28\,\au$.  According to the table assembled by Griest
\& Hu (1992), 32\% of F stars and 42\% of A stars have separations within
three Einstein radii in the first case, and  41\% of F stars 47\% of A stars
do so in the second case.  I take the average of these two cases and I assume
that 2/3 of all A and F stars have binary companions.  I then estimate
5\% of F stars and 9\% of A stars will have companions that generate
asymmetries that could be mistaken for parallax effects.  This fraction is
small, but not negligible.

	The last effect that can mimic parallax asymmetries is a binary
lens of projected separation, $b$.  If $b/r_e \sim 1$, then the light curve
generally has caustics or other features that easily distinguish it from
a standard microlensing light curve.  However, for $0.25\lsim b/r_e\lsim 0.6$
and for $1.4 \lsim b/r_e\lsim 3$, binary-lens light curves can look similar
to standard light curves except that they have a slight asymmetry.  For 
$b\lsim 0.25 r_e$ the binary acts effectively as a point lens (Gaudi \& Gould
1997b) and for $b\gsim 3 r_e$, the effect of the companion is either too
small to be noticed or causes a double-peaked rather than an asymmetric event.
It is more difficult to estimate a priori what is the level of 
contamination from
binary lenses than from binary sources because unlike the source population,
the lens population is almost completely unknown.  
However, it is possible to estimate the level of
contamination from the microlensing data themselves.

	Both the angle $\theta$ between $\bv$ and binary separation vector and 
the impact parameter $\beta$ are randomly distributed.  
At all binary separations there
are some combinations of $\theta$ and $\beta$ that give rise to easily
recognizable effects.  For example, for separations $b\sim 2 r_e$ 
and $\theta = 0$, the events will be double peaked.  For $b\lsim 0.6 r_e$,
the character of the recognizable deviations is more difficult to describe,
but can be gauged from Figure 1 of Gaudi \& Gould (1997b).  Thus, in principle
one can estimate the binary fraction and the binary-separation and mass-ratio
distributions by cataloguing these
recognizable events and determining the efficiency of their detection by
Monte Carlo simulation.  Once this distribution is known, one can predict
the number of asymmetric events that are indistinguishable from parallax
again by Monte Carlo.

	In practice, the precision of these estimates will be limited by the
small number of events that have clear characteristics of binary lenses.
However, since the rate of such clear cases is of order the rate of
asymmetric binary-lens events, this procedure should produce an estimate 
of the background that is only somewhat larger than the Poisson error.

	At present, it is possible to give only a very rough estimate of the
contamination by binary lenses.  I do so as follows.  I note that one of
the eight MACHO LMC candidates is clearly a binary.  I assume that the
number of contaminants is approximately equal to the number of clear cut 
binaries, and so estimate a contamination rate of 12\%.  Hence the overall
contamination rate from binary lenses and binary sources is about 19\%.

\chapter{Resolving the Nature of the Lenses}

	What could be accomplished with a five year followup program?  I
consider first the more modest scenario with one quarter access to a 1 m
telescope which, as discussed in \S\ 3, could be expected to yield 15 useful
measurements.  Let us suppose that most (say 80\%) of the lenses reside in
the halo and the rest are in the Galactic disk or the LMC.  Of the expected
12 halo events, only about 2/3 would have measurable parallaxes.  For
the remainder, the combination $|\cos\phi|/S$ would fall substantially below
the value $\sim 0.09$ assumed in equation \sneval.  Thus, one would expect
8 detectable events plus 1.3 background events among the remaining 7 events
with undetectable parallaxes (see \S\ 4).  (In so far as there
were Galactic-disk events, these would also be detectable but because of their
large parallaxes would usually be recognizable as such.) \ \   On the other
hand, if there were no halo lenses one would still expect 
about one additional spheroid event which would mimic the
parallax characteristics of a halo lens plus 2.7 background
events.  Assuming that the variance of the background is twice
Poisson, these two scenarios would
therefore imply $9.3 \pm 2.2$ and $3.7\pm 2.3$ asymmetric
events respectively.  These
two ranges are only marginally separated.  If there were, for example, 3 or
10 asymmetry detections, then the results would be unambiguous.  If there
were 7 detections, however, the situation would be less clear.  

	With a dedicated 1 m telescope, the two scenarios would imply
$18.6 \pm 3.1$ and $7.4 \pm 3.3$ asymmetric
events.  These ranges are well separated, and
one could unambiguously discriminate between them.

	However, even the more aggressive version of this experiment would
leave a number of questions unresolved.  For example, if a large fraction of
the events had measurable asymmetries, one could still not tell whether they
lay in a halo or in the Evans et al.\ (1998) flared disk.  Moreover, because
the measurable quantity, $K_\parallel=\cos\phi/\tilde v$, depends on the
unknown angle $\phi$, the velocity dispersion of a detected halo would be
poorly constrained.  One therefore
could not distinguish between a ``true'' ($r^{-2}$) 
halo and a heavy spheroid ($r^{-3.5}$) whose velocity dispersion would
be expected to be smaller by a factor $(2/3.5)^{1/2}\sim 0.75$
(assuming a fixed logarithmic potential).  Likewise,
one could not reliably detect rotation of the halo.  These additional pieces
of information, which are crucial to understanding the formation process
and so the nature of the halo lenses (if that is what they are) could be
obtained from satellite parallaxes, but not from the ground.

	Finally, I note that of the two bulge followup collaborations,
PLANET's telescope time is largely concentrated in the bulge season, but
GMAN has access to significant telescope time on a year round basis.  GMAN
has in fact followed several LMC events although the results have not yet
been published.  It therefore should be possible to initiate at least a 
modest version of the observing program outlined here on the basis of
currently available telescope resources.

{\bf Acknowledgements}:  I would like to thank Piotr Popowski for a careful
reading of the manuscript.
This work was supported in part by grant AST 94-20746 from the NSF and in 
part by grant NAG5-3111 from NASA.

%\bigskip
\endpage
\Ref\Alard{Alard, C., \&  Lupton, R.\ H.\ 1998, ApJ, submitted 
(astro-ph 9712287)}
\Ref\albrow{Albrow, M., et al.\ 1996, in IAU Symp.\ 173, Astrophysical
Applications of Gravitational Microlensing, ed.\ C.\ S.\ 
Kochanek \& J.\ N.\ Hewitt) (Dordrecht: Kluwer), 227}
\Ref\alb{Albrow, M.\ et al.\ 1998, ApJ, submitted}
\Ref\Alcock{Alcock, C., et al.\ 1995, ApJ, 454, L125}% parallax
\Ref\Alcock{Alcock, C., et al.\ 1996, ApJ, 463, L67}
\Ref\alc{Alcock et al.\ 1997a, ApJ, 486, 697}
\Ref\alc{Alcock et al.\ 1997b, ApJ, 490, 59}
\Ref\alc{Alcock et al.\ 1997c, ApJ, 491, L11}
\Ref\alc{Alcock et al.\ 1997d, ApJ, 491, 436}
\Ref\Ansari{Ansari, R., et al.\ 1997a, A\&A, 324, 69}
\Ref\Ansari{Ansari, R., et al.\ 1997b, A\&A, 324, 843}
%\Ref\Aubourg{Aubourg, E., et al.\ 1993, Nature, 365, 623}
\Ref\BS{Beaulieu, J.-P., \& Sackett, P.\ D.\ 1998, AJ, submitted 
(astro-ph 9710156)}
\Ref\ben{Bennett, D.\ et al.\ 1997, BAAS, 191, 8308}
\Ref\bg{Boutreux, T., \& Gould, A.\ 1996, ApJ, 462, 705}
\Ref\alc{Bennett, D.\ 1998, ApJL, submitted}
%\Ref\BT{Binney, J.\ \& Tremaine, S.\ 1987, Galactic Dynamics, (Princeton:
%Princeton Univ.\ Press)}
\Ref\bk{Buchalter, A., \& Kamionkowski, M.\ 1997, 482,782}
%\Ref\cm{Chabrier, G., \& M\'era, D.\ 1997, A\&A, 328, 83}
\Ref\evans{Evans, N.\ W., Gyuk, G., Turner, M.\ S., \& Binney, J.\ J.\ 1998, 
Nature, submitted}
%\Ref\dVF{de Vaucouleurs, G. 1957, AJ, 62, 69}
\Ref\gaudi{Gaudi, B.\ S., \& Gould, A.\ 1997a, ApJ, 477, 152}
\Ref\gaudi{Gaudi, B.\ S., \& Gould, A.\ 1997b, ApJ, 482, 83}
\Ref\gtwo{Gould, A.\ 1992, ApJ, 392, 442}
\Ref\gtwo{Gould, A.\ 1994, ApJ, 421, L75}
\Ref\gtwo{Gould, A.\ 1995a, ApJ, 441, L21}%single satellite
\Ref\gthree{Gould, A.\ 1995b, ApJ, 441, 77}
\Ref\gtwo{Gould, A.\ 1998, ApJ, 499, 000}
%\Ref\gthree{Gould, A., Bahcall, J.\ N., \& Flynn, C.\ 1996, ApJ, 465, 759}
%\Ref\gthree{Gould, A., Bahcall, J.\ N., \& Flynn, C.\ 1997, ApJ, 482, 913}
%\Ref\gthree{Gould, A., Flynn, C., \& Bahcall, J.\ N.\ 1997, ApJ, submitted
%(astro-ph 9711263)}
\Ref\gmb{Gould, A., Miralda-Escud\'e, \& Bahcall, J.\ N.\ 1994, ApJ, 423, L105}
\Ref\gtwo{Gould, A.\ \& Welch, D.\ L.\ 1996, ApJ, 464, 212}
\Ref\gh{Griest, K., \& Hu, W.\ 1992, ApJ, 397, 362}
\Ref\gtwo{Han, C.\ \& Gould, A.\ 1997, ApJ, 480, 196}
\Ref\hmc{Henry, T.\ J.\ \& McCarthy, D.\ W.\ 1993, AJ, 106, 773}
\Ref\mark{Markovi\'c, D.\ 1998, MNRAS, submitted}
\Ref\melch{Melchior, A.-L.\ 1998, A\&A, submitted (astro-ph 9712236)}
\Ref\pb{Palanque-Delabrouille, N.\ et al.\ 1998, A\&A, in press}
\Ref\pac{Paczy\'nski, B.\ 1986, ApJ, 304, 1}
\Ref\refsdal{Refsdal, S.\ 1966, MNRAS, 134, 315}
\Ref\sahua{Sahu, K.\ C.\ 1994a, Nature, 370, 275}
\Ref\tom{Tomaney, A., 1998, preprint (astro-ph/9801233)}
\Ref\tomcro{Tomaney, A., \& Crotts, A.\ P.\ S.\ 1996, AJ, 112, 2872}
\Ref\wu{Wu, X.-P.\ 1994, ApJ, 435, 66}
\Ref\zarit{Zaritsky, D., \& Lin, D.\ N.\ C. 1997, AJ, 114, 254}
\Ref\zhao{Zhao, H.\ 1998, MNRAS, submitted (astro-ph 9703097)}
\refout
\endpage
%\figout
%\endpage
\end